\documentclass[11pt,preprint]{aastex}
\shorttitle{Empirical Color-Temperature Calibration}
\shortauthors{Guy Worthey}

\begin{document}

\title{An Empirical UBVRIJHK Color-Temperature Calibration for Stars}

\author{Guy Worthey}
\affil{Department of Physics and Astronomy, Washington State
University, Pullman, WA 99164-2814}

\author{Hyun-chul Lee}
\affil{Department of Physics and Geology, The University of Texas - Pan American, Edinburg, TX 78539-2999}

\begin{abstract}

A collection of Johnson/Cousins photometry for stars with known [Fe/H]
is used to generate color-color relations that include the abundance
dependence.  Literature temperature and bolometric correction
dependences are attached to the color relations. The $JHK$ colors are
transformed to the \citet{bb88} homogenized system. The main result of
this work is the tabulation of seven colors and the $V$-band
bolometric correction as a function of $T_{eff}$, log $g$, and [Fe/H]
for $-1.06 < V-K < 10.2$ and an accompanying interpolation program.
Improvements to the present calibration would involve filling
photometry gaps, obtaining more accurate and on-system photometry,
knowing better log $g$ and [Fe/H] values, improving the statistics for
data-impoverished groups of stars such as metal-poor K dwarfs, applying small
tweaks in the processing pipeline, and obtaining better empirical
temperature and bolometric correction relations, especially for
supergiants and M stars. A way to estimate dust extinction from M
dwarf colors is pointed out.

\end{abstract}


\section{Introduction}

Stellar evolutionary tracks and isochrones calculate physical radius,
luminosity, and effective temperature. In order to compare with
observable quantities, almost always magnitudes and colors, a
transformation is essential. The need for such transformations is also
felt when integrated light models (population synthesis models) are
constructed for comparisons to colors from galaxies and star
clusters. Color-temperature transformations are also used in spectral
abundance analysis, design of observing strategies, computation of
selection effects, and a host of incidental astronomical problems.

One way to approach this problem is to calculate line-blanketed
synthetic spectra and integrate under filter transmission functions to
get fluxes, which are then zeroed by comparison with Vega or other
standard \citep{buser92,bg89}. This is a convenient
approach, but vulnerable to errors in all of the steps of the process:
incorrect atmosphere structures, incorrect or incomplete line lists,
incorrect treatment of convection, turbulence, non-LTE effects, or
line broadening, incorrect filter transmissions, inaccurate
spectrophotometry of the comparison star or stars, and finally, the
photometry of the comparison star itself. Examples of suspicious
circumstances include the fact that the blue edge of the $U$ filter is
set by the earth's atmosphere and will inevitably change with time and
place, the fact that half the lines in the solar spectrum have yet to
be identified \citep{k92a}, and the fact that the absolute flux
calibration for stars is uncertain by about 5\% [e.g. \citet{berri}]
Recent examples of synthetic calibrations are \citet{vc03},
\citet{vaz96}, \citet{lej98}, and \citet{houd00}.

There is need for empirical alternatives in the literature, and the
present paper attempts to fill in that gap somewhat.  The inspiration
for this work comes from \citet{green88}. Green describes a global
color-$T_{eff}$ calibration generated for attachment to the Revised
Yale Isochrones \citep{ryi} that provides colors tabulated for a
(long) list of temperatures, surface gravities, and [Fe/H] values. The
strategy used by Green was to begin with empirical color-color
relations for solar-metallicity stars, and adopt the ridge line as the
starting place. Then one attaches a color-$T_{eff}$ relation and adds
[Fe/H] and gravity dependence by working differentially within
synthetic color tables. The approach here is similar, but stays in the
empirical regime longer in that the gravity and abundance dependences
are fit to the stars themselves rather than via synthetic
photometry. In a second phase, $T_{eff}$ and the bolometric
corrections are attached to the fitted multidimensional space of $V-K$
color, gravity, and abundance. Synthetic colors are used at very low
weight to guide the fits where there are few or no stars in the
sample, but seemed to be superfluous except for metal-poor M giants,
which do not exist in nature. Only oxygen-rich stars are considered
here. Color-temperature relations for carbon-rich giants are given in
\citet{bergeat01}.

This paper is divided into sections on procedure, literature
comparisons, and a concluding section. Supporting material
(color-temperature table and interpolation program) is available at
http://astro.wsu.edu/models/.

\section{Procedure}
\subsection{Stellar Data}

The nucleus of the photometry catalog is the compilation of
\citet{mm78}, which is firmly Johnson-system. Many other photometry
sources were included.  These include \citet{veed74},
\citet{bessell91}, \citet{stet81}, \citet{daC90}, \citet{wbf83},
\citet{wb83}, \citet{braun98}, \citet{c83}, \citet{cohen78},
\citet{elias82}, \citet{elias85}, \citet{f78}, \citet{f79},
\citet{p80}, \citet{cohen80}, \citet{f81}, \citet{dac81},
\citet{cf82}, \citet{f83}, \citet{fmb90}, \citet{leggett},
\citet{dahn}, and the 2MASS point source catalog. 

As a first try at assigning abundance measurements to the list of 4496
stars, the \citet{cay01}, \citet{mcw}, and \citet{edv93} abundance
catalogs were consulted. M giants with good photometry were
artificially assigned an [Fe/H] of zero except those that belong to
clusters, in which case the cluster metallicity was adopted. M dwarfs
were assigned an [Fe/H] based on their kinematics, most of which came
from \citet{veed74}. Young disk objects were assigned [Fe/H] $= -0.1$,
old disk $-0.5$, and halo $-1.5$. Cluster stars naturally inherited
the cluster metallicity. Cluster abundances came from mostly
secondhand compilations \citep{wor94a, braun98, f78}. An abundance of
+0.3 was adopted for NGC 6791 \citep{wj03}. LMC field stars were
assigned $-0.3$ and stars in the SMC $-0.6$. Some supergiants and very
hot stars were artificially assigned [Fe/H] $=0$ when no abundance was
available, but many had literature abundances. Unfortunately, a
complete citing of the abundance sources cannot be given, as notes on
some of the (perhaps 5\%) abundance assignments have been lost. A
total of 2090 useable stars had abundances, although the number is
considerably less for any given photometric color. Odd holes appear in
the final data set. For instance, a primary source for M dwarf
colors is \citet{veed74} from which $J$-band data is missing. $U$-band
data is hard to find for cool stars. Available $R$-band data has gaps
as well. To try to fill in the ``K dwarf desert'' (see below) we also
scoured \citet{nstars1,nstars2,casagrande} for photometry and
abundance information.

Solar metallicity mean relations for all spectral types from
\citet{johnson66} and \citet{bb88} were included in the list, with
[Fe/H] $= -0.1$ assumed for these ``stars''.

\subsection{Photometric Systems}

All, we think, would agree that the collection of the various
photometric systems are, collectively, an admirable effort but also a
bit of a mess due to the fact that one telescope/site/detector
combination is a unique thing, not transferable to other telescopes in
other places with different equipment. This is mostly overcome by
observing standard stars that have been observed many times by one
setup and should therefore be internally homogeneous: a photometric
``system.''  The \citet{mm78} catalog is on the ``Johnson''
photometric system.  For colors involving $RI$, the target system was
``Cousins'' and we applied the tranformation equations of
\citet{bessell79} and \citet{bessell83} to transform the Johnson data
except for $R-I$, for which we used a tracing of Figure 3 from
\citet{bessell83} rather than the formula given in the
paper. Additional optical data that was already on the Cousins system
was left there.

Infrared data was imported from 5 different systems (and this is mild
compared to the number of systems that have proliferated over the
years). As a target system, we chose the homogenized system of
\citet{bb88}. Transformations from Johnson-system, CIT-system, and
AAO-system were used as provided in \citet{bb88}. Some 2MASS data,
mostly attached to NGC 6791 stars in the present stellar catalog, were
tranformed via \citet{carp01} formulae to the \citet{bb88} system.

Corrections for interstellar extinction were done using the
\citet{car89} extinction curve. Note that corrections are applied
differently for Johnson $RI$ than for Cousins $RI$ since the filters
are at substantially different wavelengths. Such wavelength differences
cause negligible correction differences at infrared wavelengths. 

\subsection{Color-color Fitting}

The photometrically-homogenized, dereddened stellar data were then
presented to a series of additional processing steps. A multivariate
polynomial fitting program (a modification of the one used in
\citet{wor94a} to fit spectral indices as a function of stellar
atmosphere parameters) was applied to the data. The dependent variable
chosen was $V-K$ because it is monotonically increasing with
temperature and insensitive to abundance. $V-K$ is a fabulous
temperature indicator in stars cooler than the sun, and, because of
its monotonicity, can still serve as a temperature-like variable for
hotter stars.  Terms of up to order $(V-K)^6$ could be included, and
up to quadratic terms of log $g$, [Fe/H], and cross-terms.  Chemically
peculiar stars such as carbon stars were excluded from the fit.  

\begin{figure}
\plotone{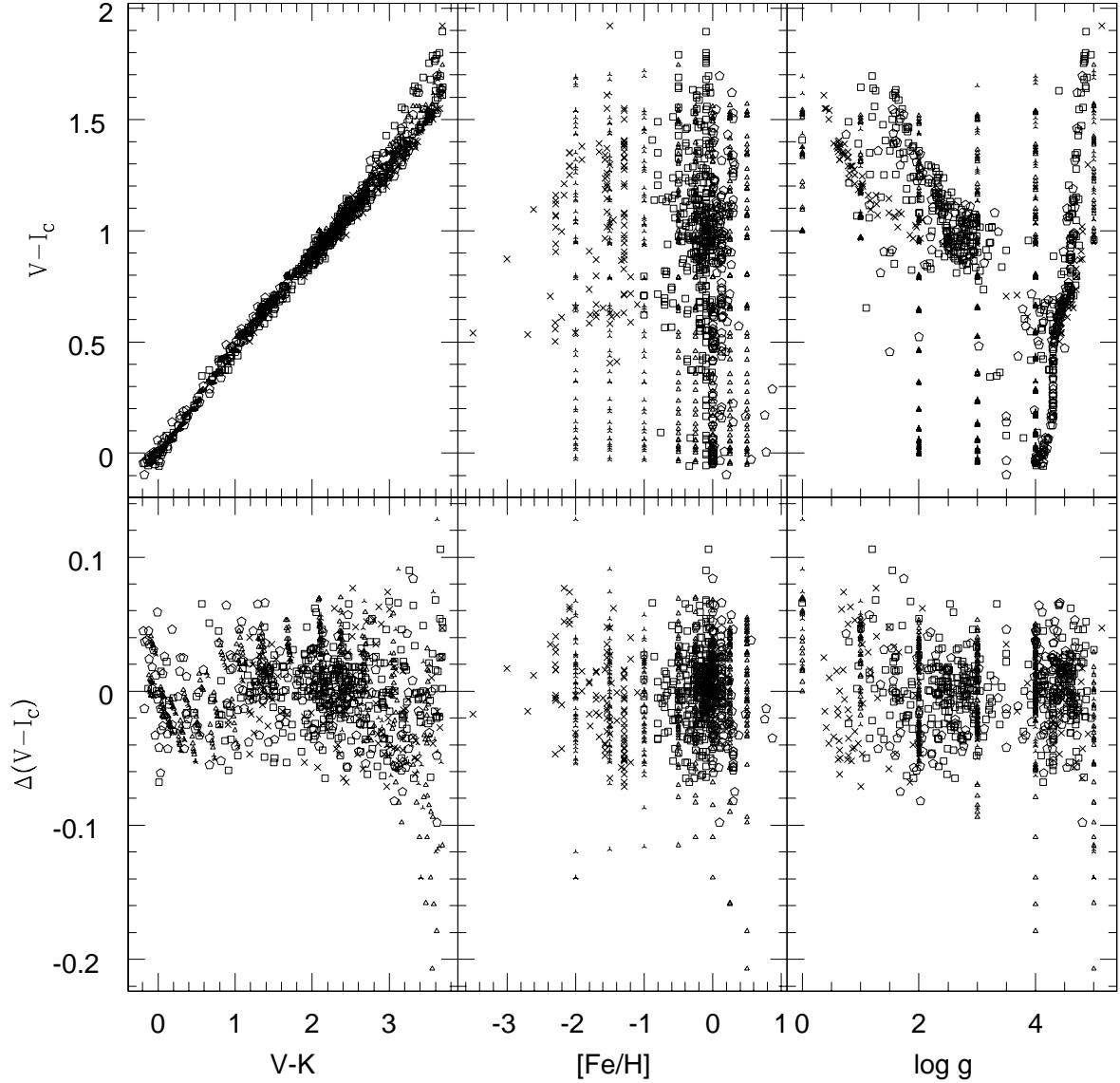}
\caption{An illustration of the fitting process in the warm
temperature range, using the $V-I$ color. Smaller-size 3-vertex
symbols are synthetic photometric points that were not allowed to
affect the fit if real stars were present, open for metal-rich
``stars'' and skeletal for metal-poor ``stars.'' Open pentagons are
metal-rich stars, open squares are between [Fe/H] = $-1$ and 0, and
skeletal squares are metal-poor. These choices can be directly seen in
the two, middle [Fe/H] panels. The top row of panels is $V-I$ versus
$V-K$, [Fe/H], and log g, and the bottom row of panels is the data
minus fit residuals versus the same three variables. These plots
vaguely mimic what the fitting program shows as it operates, although
the fitting program can better isolate and display arbitrarily defined
stellar groups, and also shows fits.
\label{fig:6panel} 
}
\end{figure}

The color range was divided into 5 widely overlapping sections and
each range was independently fit. For example, the second-hottest temperature
section, for the color $V-I$, is displayed in Figure \ref{fig:6panel}.
The specific polynomial terms allowed in the fit could be different
for each temperature section. This allowed, for instance, [Fe/H]
sensitivity to be manually phased out if desired. The fits were done
many times. Outlier data points were rejected manually with the aid of
a graphical interface that allowed the name and parameters of each
star to be scrutinized before rejection. Before one (of seven) color
fits in one (of five) temperature regimes passed inspection, it was
examined, both raw and as residuals from the fit, against all three
variables of color, gravity, and abundance. Data rejection and
polynomial term additions and subtractions were done iteratively with
the aid of f-test statistics. In an approximation of what appears
during the fitting process, Figure \ref{fig:6panel} shows both raw
data and residuals after the fit as a function of $V-K$ color, [Fe/H],
and log $g$, with symbol types varying as a function of
abundance. Synthetic color points are also shown, for purposes of
illustration, although we emphasize that the synthetic colors did not
influence the fits except for stars that do not exist in nature.

\begin{figure}
\plotone{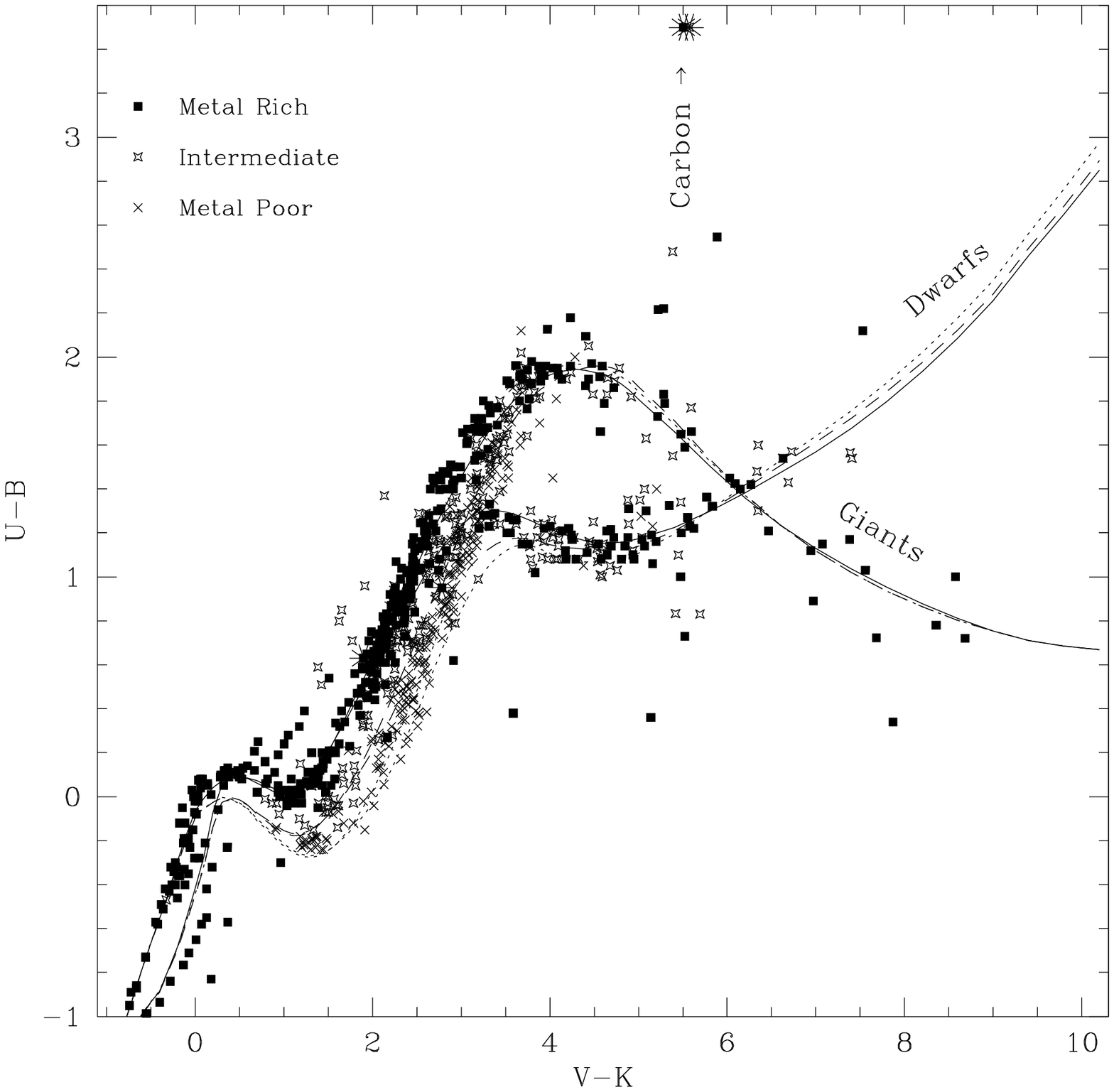}
\caption{The $U-B$, $V-K$ color-color diagram for unculled stars.
Stars have different symbol types for metal rich ([Fe/H] $> -0.2$),
metal-poor ([Fe/H] $< -1.2$), and intermediate abundance
ranges. Calibrations for typical giant and typical dwarf gravities are
drawn in solid for [Fe/H] $= 0$, dashed for [Fe/H] $= -1$, and dotted
for [Fe/H] $= -2$. Most carbon stars (asterisks) are not plotted as
they stretch beyond the plot limits along a line from the plotted ones up
to $(V-K,U-B)\approx(6,6)$. 
\label{fig:ub1} 
}
\end{figure}

\begin{figure}
\plotone{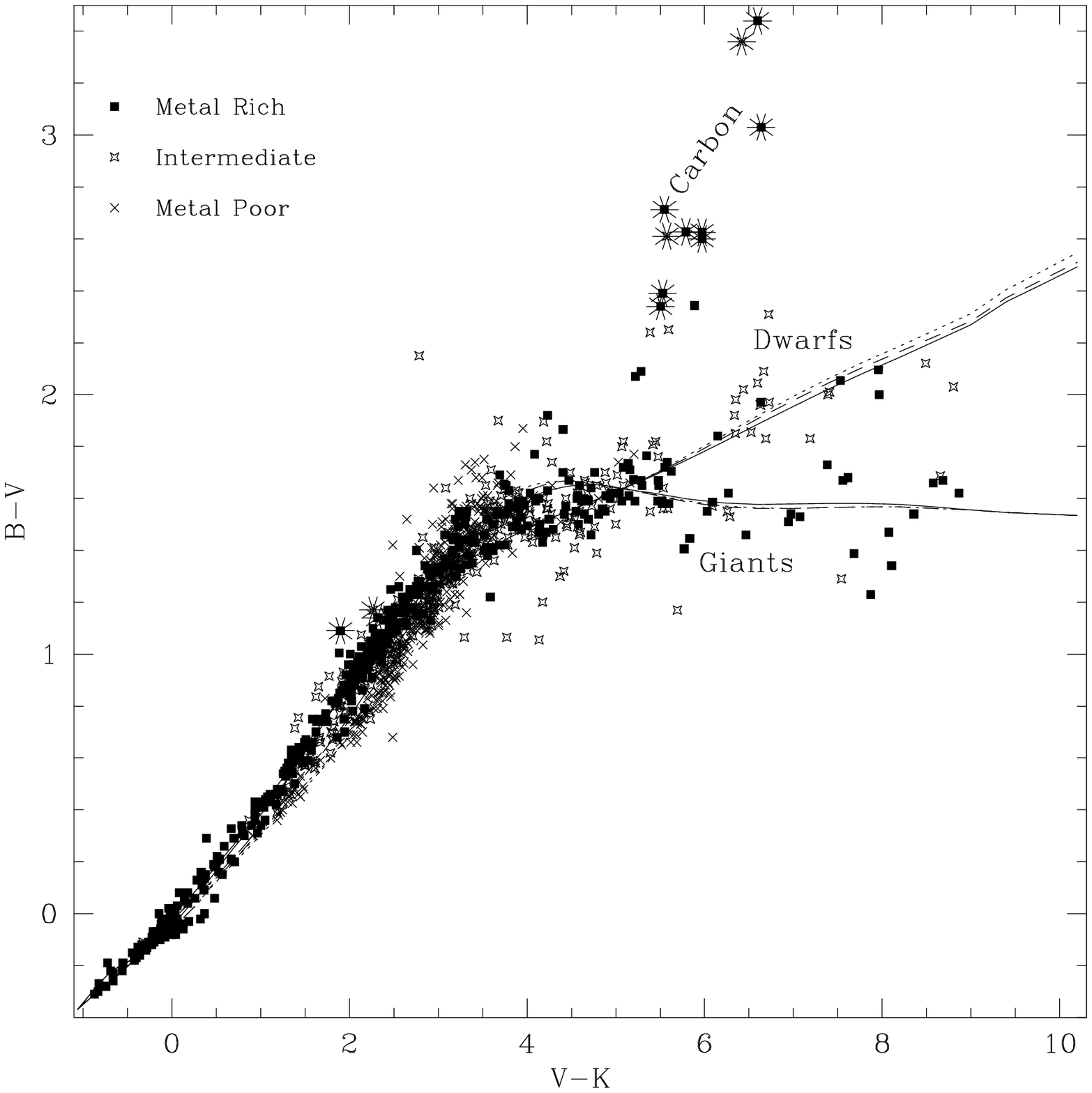}
\caption{The $B-V$, $V-K$ color-color diagram for unculled stars.
Stars have different symbol types for metal rich ([Fe/H] $> -0.2$),
metal-poor ([Fe/H] $< -1.2$), and intermediate abundance
ranges. Calibrations for typical giant and typical dwarf gravities are
drawn in solid for [Fe/H] $= 0$, dashed for [Fe/H] $= -1$, and dotted
for [Fe/H] $= -2$. Carbon stars are shown as asterisks.
\label{fig:bv1} 
}
\end{figure}

\begin{figure}
\plotone{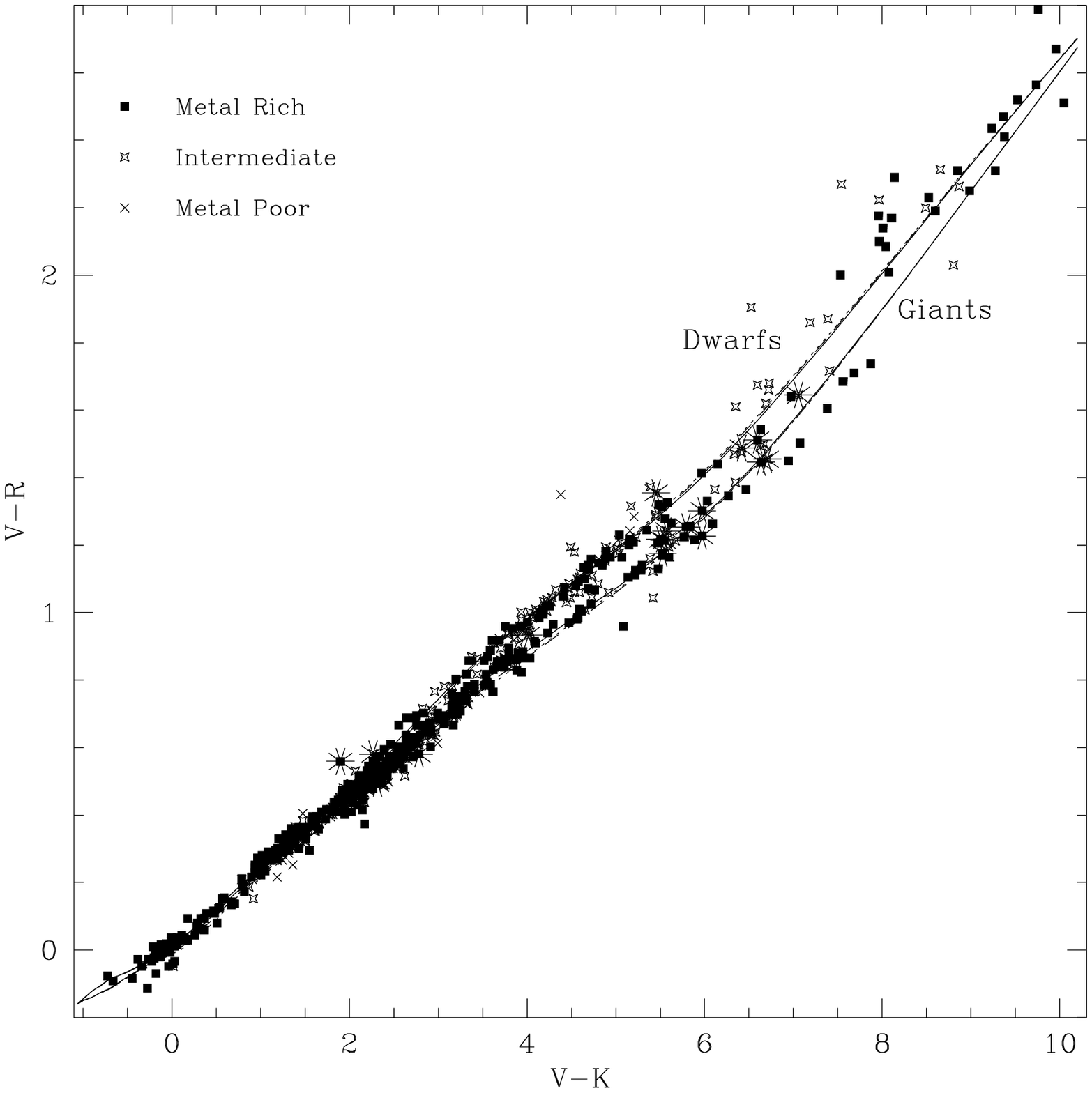}
\caption{The $V-R$, $V-K$ color-color diagram for unculled stars.
Symbols and line styles are as in Figure \ref{fig:bv1}.
\label{fig:vr1} 
}
\end{figure}

\begin{figure}
\plotone{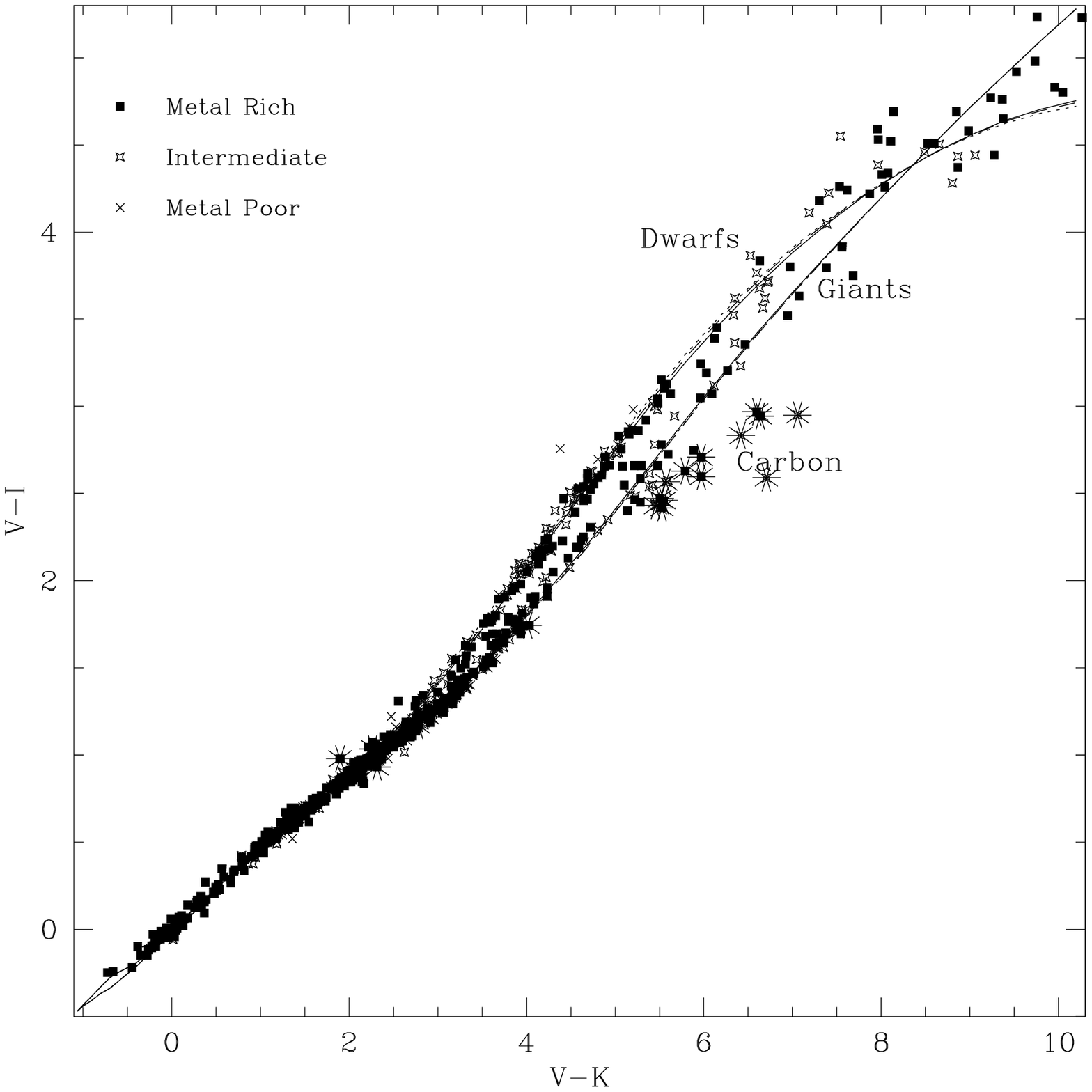}
\caption{The $V-I$, $V-K$ color-color diagram for unculled stars.
Symbols and line styles are as in Figure \ref{fig:bv1}.
\label{fig:vi1} 
}
\end{figure}

\begin{figure}
\plotone{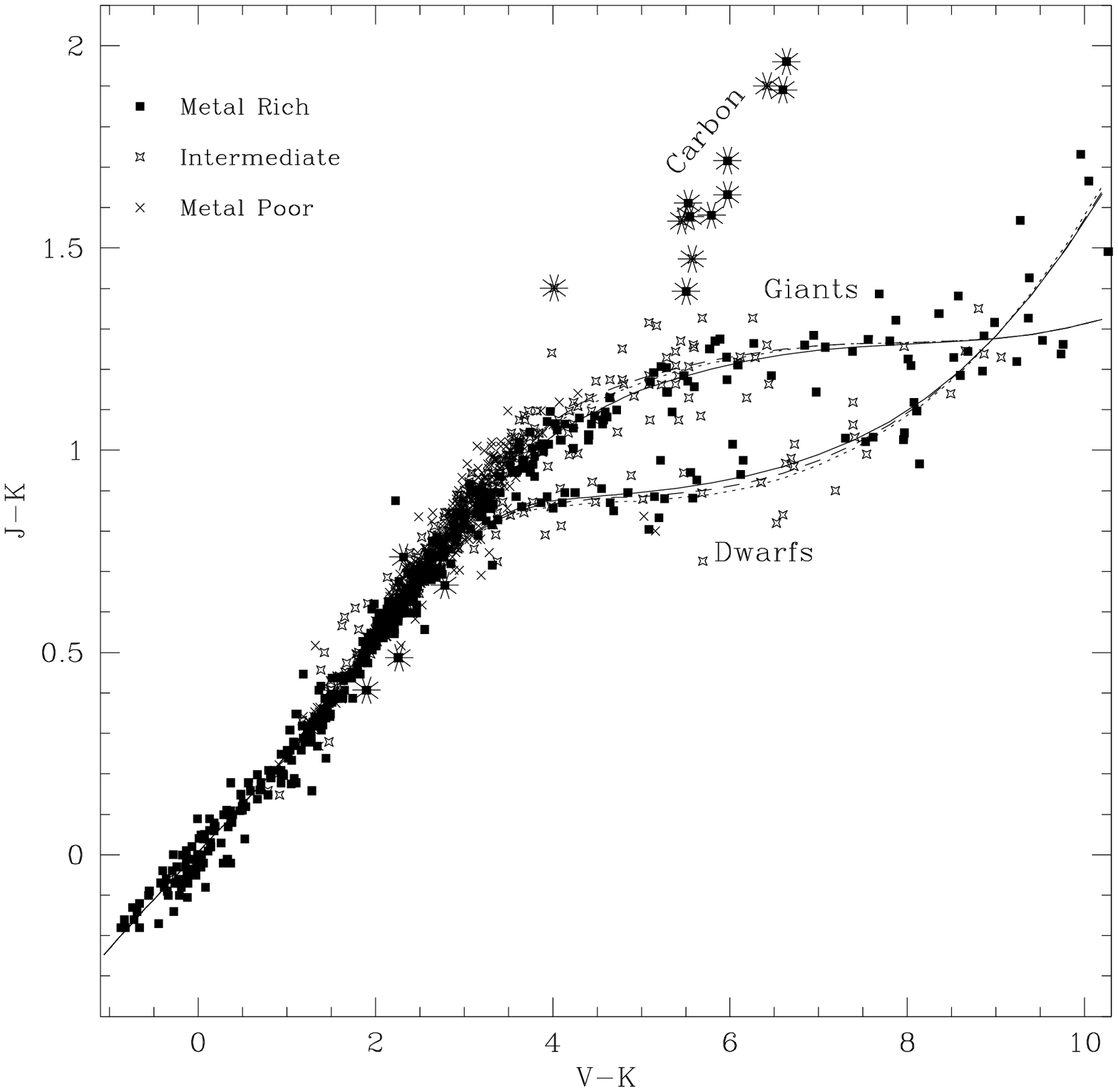}
\caption{The $J-K$, $V-K$ color-color diagram for unculled stars.
Symbols and line styles are as in Figure \ref{fig:bv1}.
\label{fig:jk1} 
}
\end{figure}

\begin{figure}
\plotone{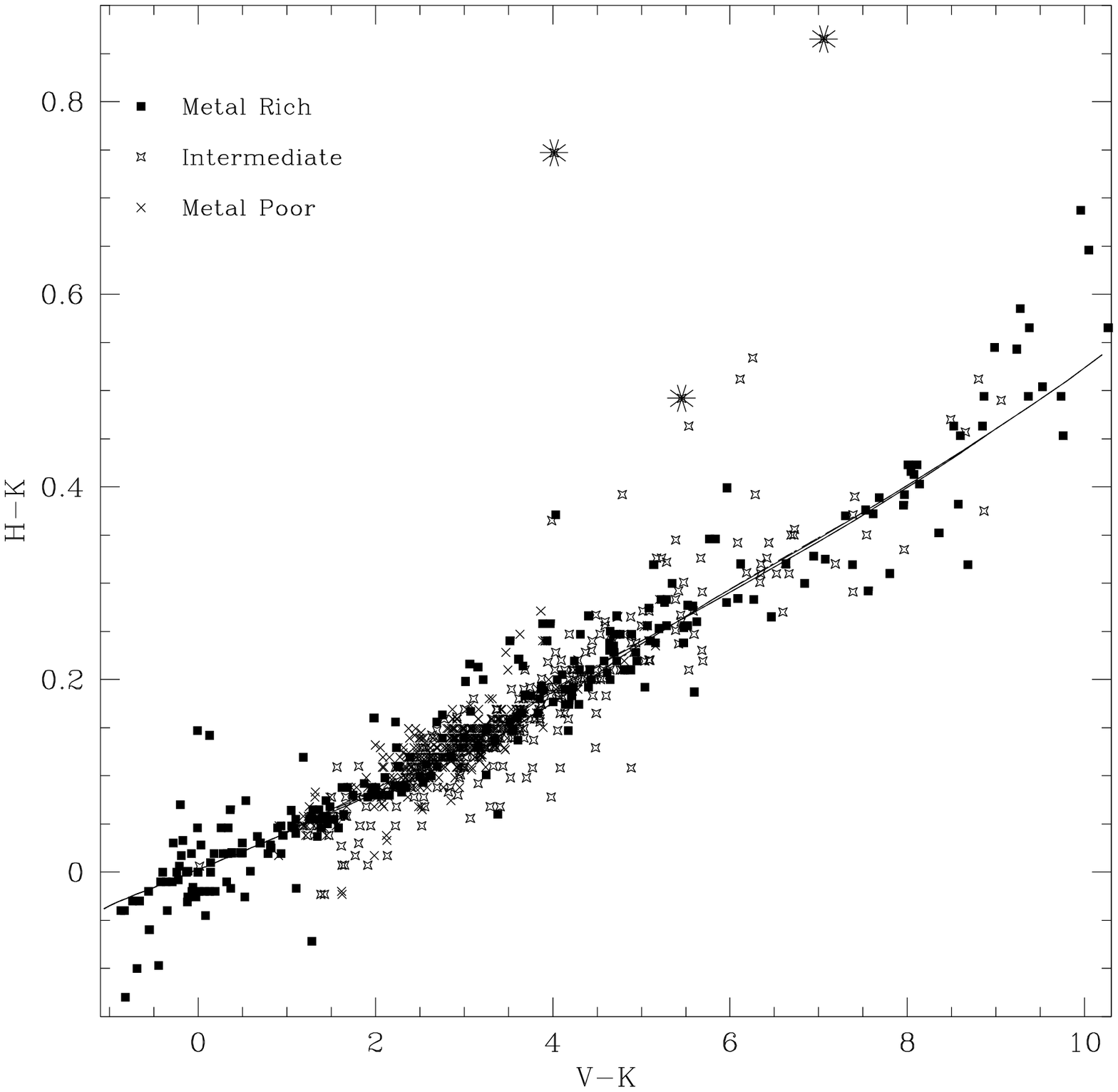}
\caption{The $H-K$, $V-K$ color-color diagram for unculled stars.
Symbols and line styles are as in Figure \ref{fig:bv1}.
\label{fig:hk1} 
}
\end{figure}

The final polynomials were combined in tabular form, using a weighted-mean
scheme wherein the middle of each $V-K$ section was weighted strongly
compared to the edges of each section. [Fe/H] and log $g$ were
tabulated in 0.5 dex intervals, $-2.5 \leq$ [Fe/H] $\leq 0.5$, and $-0.5 \leq$
log $g \leq 5.5$.  The resultant color-color relations are illustrated in
Figures \ref{fig:ub1}, \ref{fig:bv1}, \ref{fig:vr1}, \ref{fig:vi1},
\ref{fig:jk1}, and \ref{fig:hk1}. All stars, even if they were
rejected during the fitting process, are included in the
figures. Carbon stars are included, but only for illustrative
purposes; fits were not attempted and one is again referred to the
work of \citet{bergeat01}.

For very hot stars of O and B spectral type an additional color-color
table that was crafted by hand from either summary color-spectral type
relations or from our own color-color relations was employed
to refine the color-color relations for the hottest
stars. The sources consulted were \citet{k82,toku,vacca}, and our own
color-color plots. There is basically no abundance leverage for very
hot stars, so we assumed a zero metallicity dependence. These tabulated
average values were included in the polynomial fits as if they were
individual stars. 

In the hot star regime, two uncertain areas came to light that deserve
mention as regards dwarf vs. supergiant colors. First, in $U-B$,
\citet{k82} data imply a large and distinct color separation between
dwarfs and supergiants, but the (few) supergiants available in our
list did not follow the literature trend. Nor were the polynomials
flexible enough to track these changes, mostly because, for O stars,
the difference in surface gravity is very minor [4.15 (dwarfs)
vs. 4.09 (supergiants) according to \citet{vacca}]. In the end, we
performed a weighted average between the polynomial fits and the
tabulated values. There is probably considerable uncertainty left in
the supergiant $U-B$ colors, perhaps several tenths of a
magnitude. This is one area that could be vastly improved with more
photometry, with the caveats that reddening is often a huge factor for
these intrinsically bright, usually distant stars and the fact that
fast rotation introduces an inclination angle dependence in the
colors. Users wishing to avoid this entirely may want to feed our
interpolation program artificially high gravities for stars hotter
than about 9000 K. The second area of debate was that the tabulated
$H-K$ colors of \citet{toku} for O supergiants were about 0.09 mag
redder than for O dwarfs. In this case, we saw no trace of such a
trend in our data: a few stars were that red, but they were all
dwarfs. We allowed the polynomial fit (which, in that regime, was a
function of temperature alone) to determine the final color-color
relation. In the middle of the temperature range, a small gravity
dependence was indicated, but no dependence on [Fe/H] was ever
statistically significant.

In the regime of cool giants, there is a strong evolutionary effect
such that metal-poor stellar populations do not generate M-type
giants. The rich globular cluster 47 Tucanae is on the cusp,
containing 4 long-period variable stars at the tip of its giant branch
at [Fe/H] $\approx -0.8$. The SMC, at present-day [Fe/H] $\approx
-0.6$, generates some M and Carbon stars, but mostly because of
intermediate-age populations that grow very bright (and cool)
asymptotic giant branches. Thus, there is a sharp transition from
excellent metallicity coverage for K giants to very limited
metallicity leverage for M giants, exacerbated by the fact that M
giant abundances are hard to measure. In M dwarfs, where stars of all
metallicities exist in our list, there is an interesting, strong
convergence of color-color sequences as a function of metallicity so
that G dwarfs have a very strong [M/H] dependence, there is a
transition in K dwarfs, and M dwarf colors have no detectable [M/H]
dependence. In fitting, therefore, the [M/H] dependence was gradually
removed for cooler and cooler stars, for the giants because cool,
metal-poor stars do not exist, and for the dwarfs because the [M/H]
dependence removes itself empirically.

\subsection{Temperatures}

\begin{figure}
\plotone{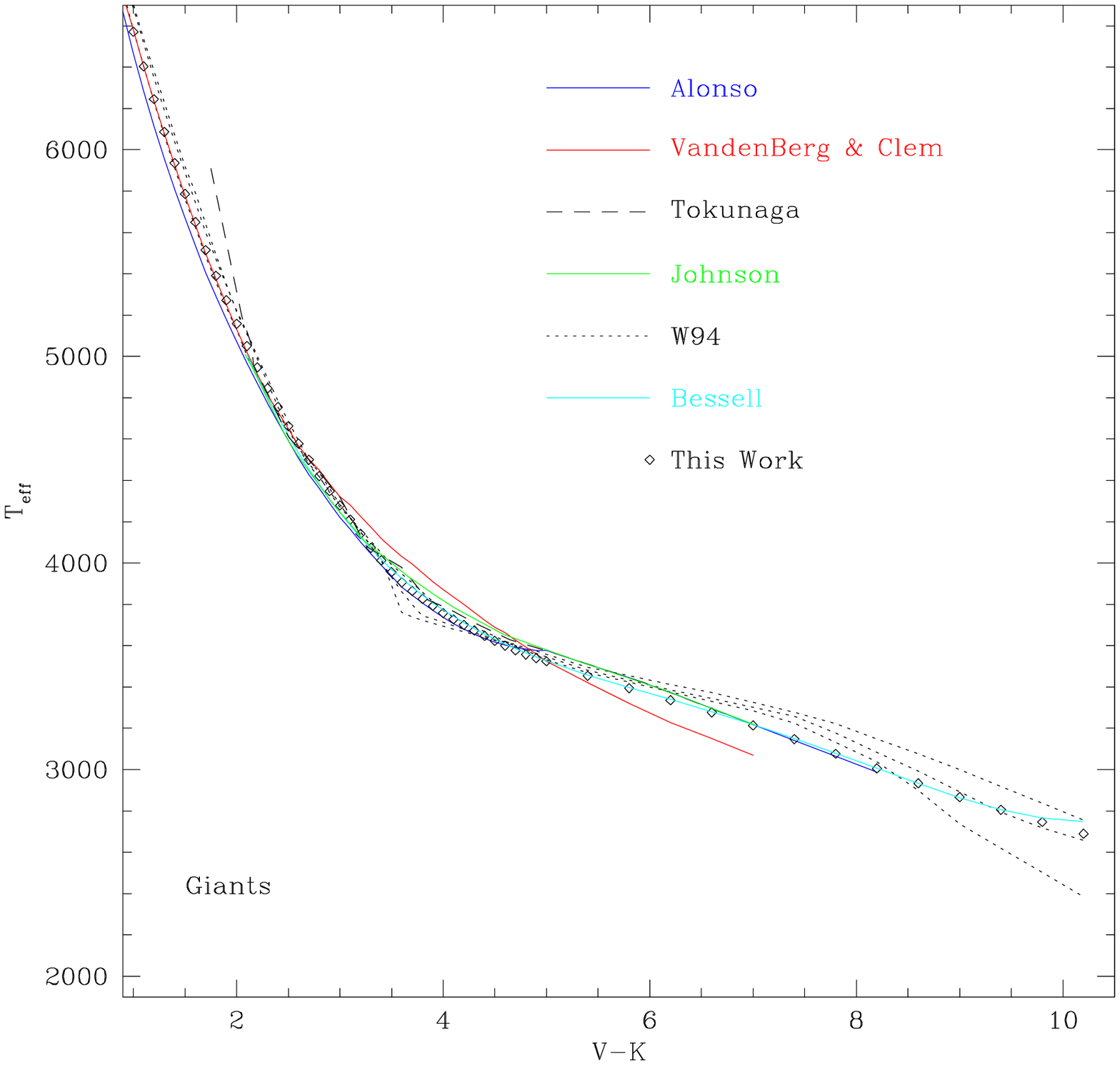}
\caption{Temperature-$V-K$ calibrations for cool, solar abundance
giants.  Lines are color-coded for the calibrations of
\citet{alonso99a,alonso99b,vc03,toku,johnson66,wor94b,bessell98}.  Our
adopted relation is shown as diamonds.
\label{fig:mash2gt} 
}
\end{figure}

\begin{figure}
\plotone{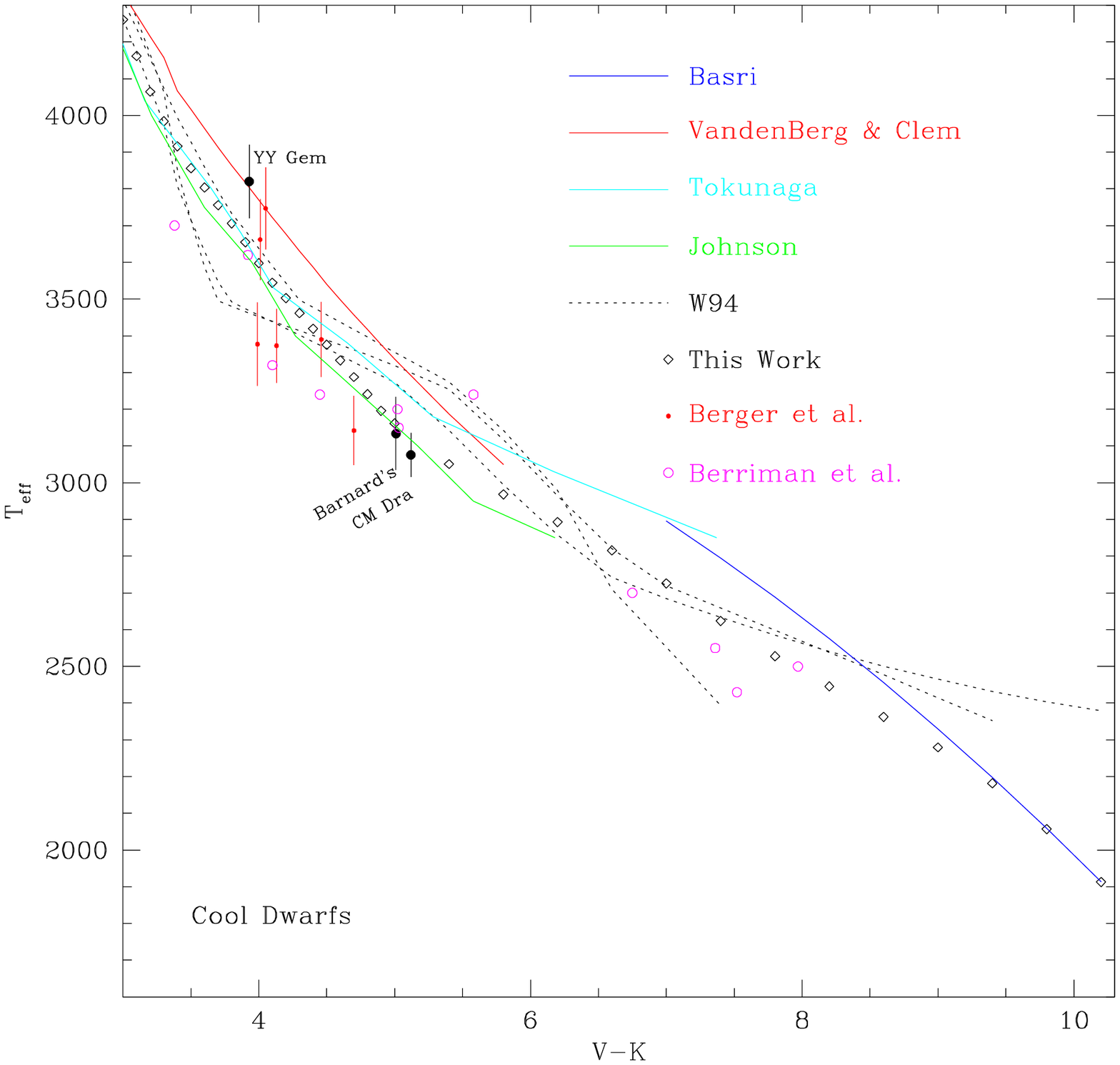}
\caption{Temperature-$V-K$ calibrations for cool, solar abundance dwarfs.
Lines are color-coded for the calibrations of
\citet{vc03,toku,johnson66,wor94b,bessell98} and
\citet{basri00}. Red dots with error bars are M dwarfs are from
\citet{berger06} and magenta open circles are TIRFM temperatures and
photometry from \citet{berri}. YY Geminorum's temperature is from
\citet{torres}, Barnard's star from 
\citet{dawson04}, and CM Draconis's from \citet{viti}.
Our adopted relation is shown as diamonds.
\label{fig:mash2dw} 
}
\end{figure}

Due to our approach of fitting color-color relations internally as a
function of gravity and abundance, attachment of temperature scales
could, in principle, be done for any color-temperature relation in any
part of the parameter space. The first iteration of this process was
to layer color-temperature relations on top of each other until the
whole parameter range was covered, and to take the median in regions
where more than one relation applied. This is illustrated in Figs. \ref{fig:mash2gt} and \ref{fig:mash2dw}. For FGK giants \citet{alonso99a}
and \citet{alonso99b} were used. These works include a specific [Fe/H]
dependence, and the average of both $V-I$ and $V-K$ relations were
used. For appropriate runs of temperatures and gravities, \citet{vc03}
$V-I$ was translated to $V-K$ via our color-color relations. In a
similar manner, the synthetic fluxes of \citet{k92} and
\citet{bbsw,bbsw2} fluxes were combined and translated to colors as in
\citet{wor94b}. In this case, both $V-R$ and $V-I$ were translated to
$V-K$ via the emprical color-color relations and plotted along with
the untweaked $V-K$ - $T_{eff}$ relations. \citet{toku} developed
average color-temperature relations for sequences of supergiants,
giants, and dwarfs using literature temperature
scales. \citet{bessell98} gives color-temperature sequences for
solar-abundance dwarfs and giants based on different model
atmospheres, and we also referred to the empirical cool giant
sequences of \citet{rid80,dyck96}. The color-temperature sequences of
\citet{johnson66} are also included. For the coolest dwarfs, analysis
of the data of \citet{basri00} yielded a relation as a function of
$I-K$ color, specifically $T_{eff} = -460.25\times (I-K) + 4323$,
valid for $I-K > 2.9$. Adopting this relation meant that the final
temperature assignments for the coolest dwarfs needed to wait for the
final color-relations to be fixed. Given the disparate ingredients, the final adopted temperatures were hand-guided a fair amount.

For example, M dwarfs with known angular diameters, but not separately summarized in existing color-$T_{eff}$ calibrations, were also included in the
mix. Eclipsing binaries YY Geminorum \citep{torres} and CM Draconis
\citep{viti} were supplemented with interferometrically-derived
temperatures from \citet{berger06} and, in the case of Barnard's star,
from \citet{dawson04}. $VK$ photometry came from either our own
catalog or that of \citet{leggett}.  The temperature estimates of
\citet{berri} for eleven dwarfs are also plotted. These are more
indirect temperature estimates from the ratio of the bolometric flux
to the flux at an infrared wavelength, the total to infrared flux
ratio method (TIRFM).  The position of especially the cooler stars was
influential in our adopting a somewhat cooler temperature scale around
3000 K than the bulk of the published calibrations. 

The fits are good to a limit of $V-K=10.2$. Since cool dwarfs and
giants have different temperature scales, this corresponds to
approximately $T_{eff}=2700$ K for solar-metallicity giants and
$T_{eff}=1914$ K for solar-metallicity dwarfs. Not that it proves or
illustrates anything significant, but the sun's $B-V$ comes out to be
0.66 mag in the final calibration, which compares well with literature estimates \citep{taylor, gray1, gray2}.

\subsection{Bolometric Corrections}

\begin{figure}
\plotone{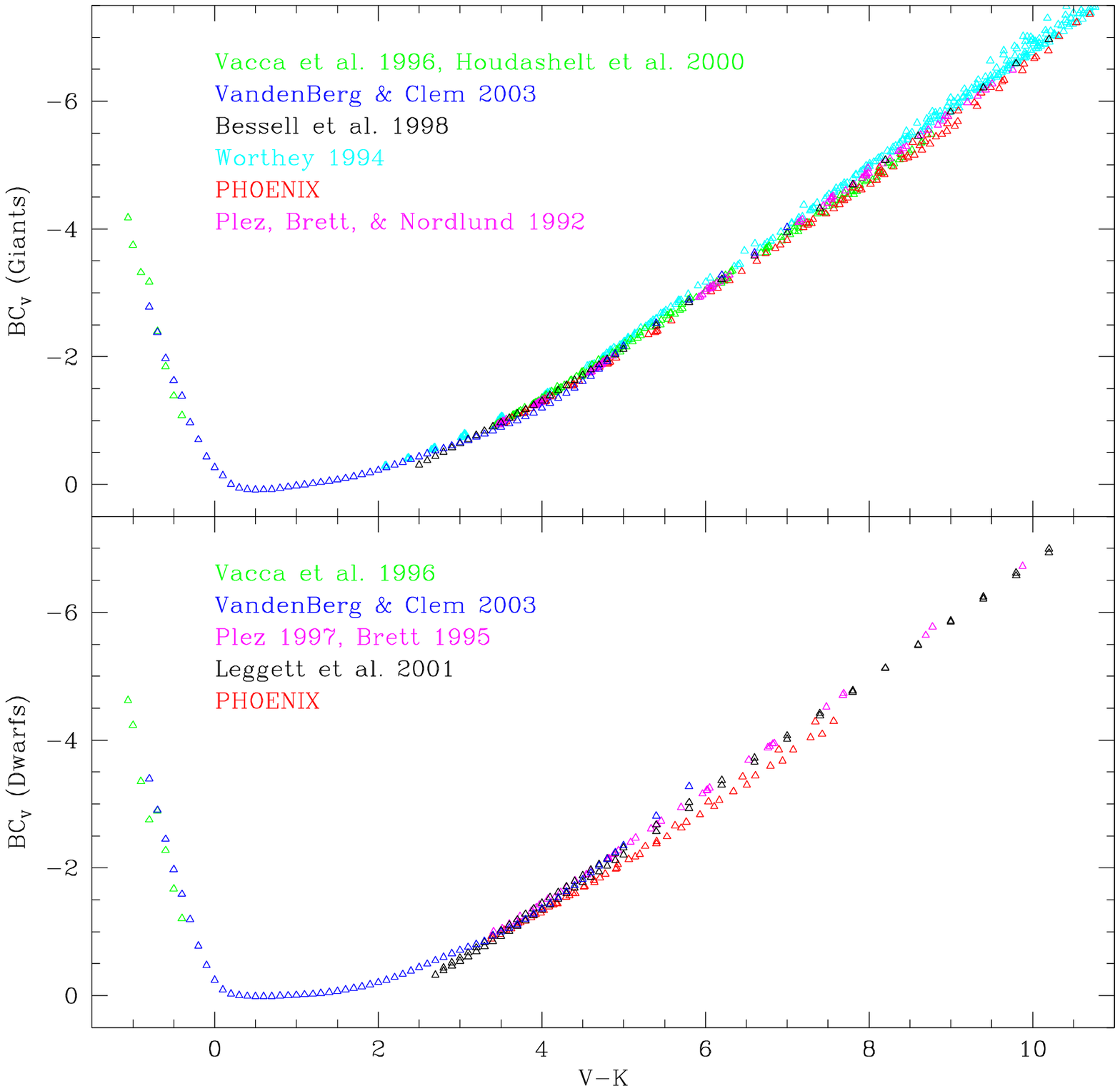}
\caption{$V$-band bolometric corrections for dwarfs and giants.
The top panel is a sequence of giants, and the bottom panel is a
sequence of dwarfs, both for near-solar metallicity. The symbol key is
marked on the plot itself with sources
\citet{vacca} for hot stars, \citet{houd00} for cool stars, \citet{vc03,bessell98,wor94b,plez92,brett95}, and \citet{leggett},
plus ``PHOENIX,'' which refers to fluxes produced from the Phoenix
code \citep{allard95} with colors generated as in \citet{wor94b} and 
``Plez 1997,'' which refers to a private communication that was subsequently
published in \citet{bessell98}. 
\label{fig:bc} 
}
\end{figure}

The last item to be added was the $V$-band bolometric correction
(BC). Since they were that last item in the chain, BCs could be
inserted as a function of color or of temperature and for any
passband. As for temperature scales, a variety of empirical and
theoretical options were intercompared. The \citet{vc03} BCs were
adopted for the middle of the temperature range, supplemented by the
\citet{vacca} formula for $4.40 < {\rm log}\ T < 4.75$ for the hottest
dwarfs and supergiants. \citet{vc03} have a solar BC$_V = -0.09$ mag,
and other scales were zero point adjusted to match. The \citet{wor94a}
BCs needed a 0.03 mag shift to match that, for example. At the cool
end, for both giants and dwarfs, the \citet{vc03} BCs drift slightly
from most calibrations, as seen in Figure \ref{fig:bc}. For giants, we
adopted the average, empirical-plus-theoretical $K$-band BC from
\citet{bessell98}, read from their Figure 20. For cool dwarfs, we
adopt the $K$-band (UKIRT IRCAM3 system) BCs of \citet{leggett}. We
extended their polynomial slightly to reach our $V-K=10.2$ cool limit.
One subtlety regarding the \citet{leggett} calibration should be
mentioned. They give two polynomial fits to the $K$-band BC, one as a
function of $I-K$ and one as a function of $J-K$. We adopt the $I-K$
version, as the $J-K$ version drifts significantly from the $I-K$
version at warmer temperatures. The cause of this drift is increased
scatter in the $J-K$ diagram, or, more fundamentally, the fact that
both $J$- and $K$-bands are on the red tail of the blackbody curve for
the warm half of the temperature range covered, so that $J-K$ as a
temperature indicator has a small temperature range per unit error.

The main calibrations employed are plotted in Figure \ref{fig:bc},
along with citations. For clarity, the fitted result and also the BCs
of \citet{buzzoni} are omitted. Note that, plotted as a function of color, and
as predicted by synthetic fluxes, the bolometric corrections are a
very weak function of abundance and gravity. This is a
degeneracy. That is, increasing a cool giant's abundance (for example)
will make it redder and give it a larger (absolute value of the)
$V$-band BC. Such vectors lie closely along the trend caused by
temperature, so BCs are strongly covariant with $T$, log $g$, and
[M/H] when plotted versus $V-K$. We exploit this for cool stars by
adopting BCs that vary as a function of color alone. Gravity and
abundance dependence then is inherited from the gravity and abundance
variations of the color-color diagrams. The various relations were
combined via temperature dependent weighted means, where the weights
were chosen to de-emphasize outliers.

Table 3, the full-length version of which is given in the
electronic version of this journal, gives the final calibration in
grid form. An ASCII version, interpolation program, and other
supporting material is available at http://astro.wsu.edu/models/.


\section{Comparisons and Discussion}

The wealth of comparison data that we could be checking against is too
vast to illustrate completely in the pages of this journal, so we
limit ourselves to a few key examples. 

\subsection{Cool Regime}

\begin{figure}
\plotone{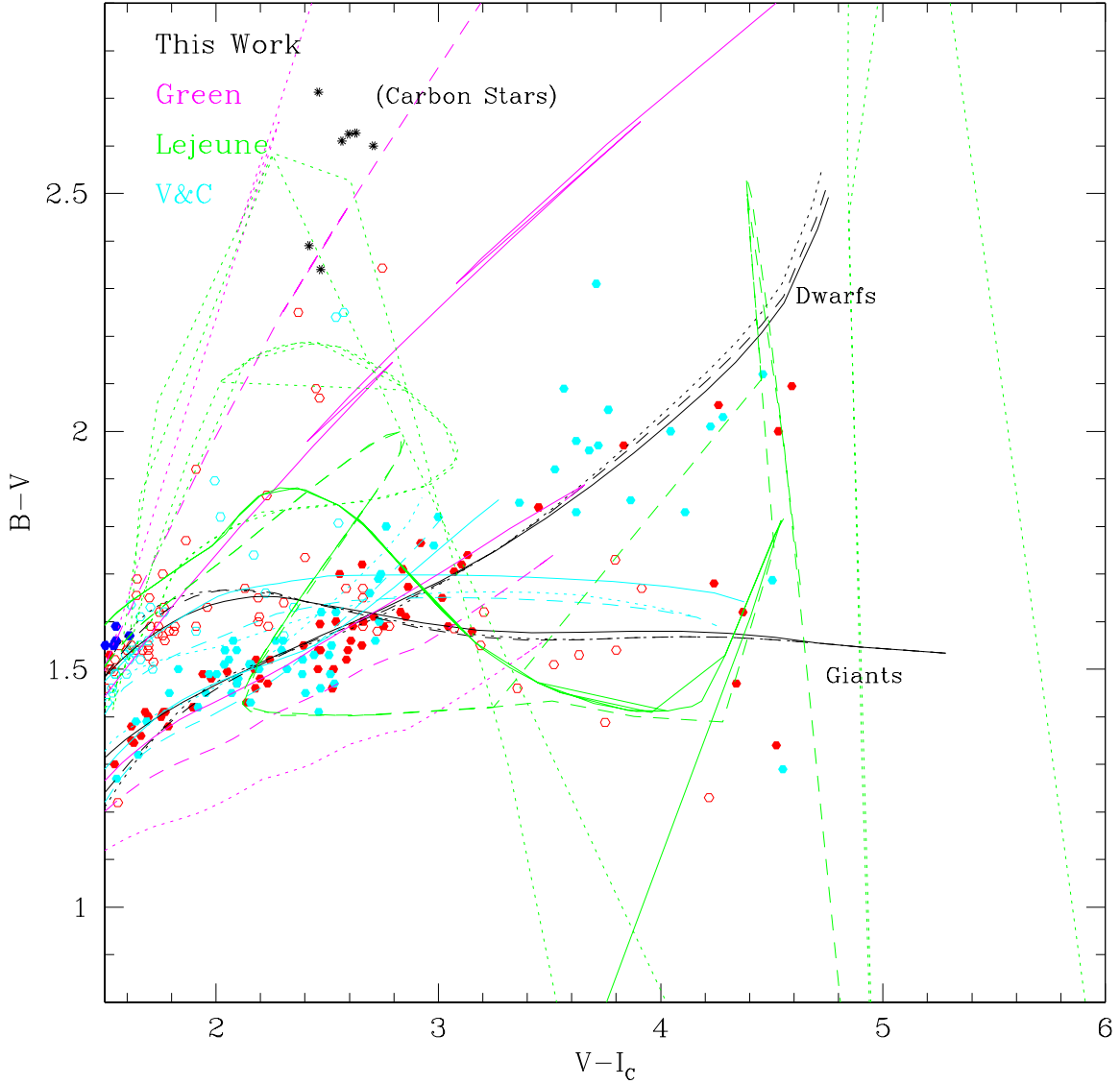}
\caption{The $B-V$, $V-I$ color-color diagram for M stars. Red dots
are stars with [Fe/H] $> -0.2$ and blue dots are stars with [Fe/H] $<
-1.2$ with intermediate stars cyan. Giants are open symbols, which
dwarfs are filled. The data are unculled. Calibrations for giant and
dwarf color-color 
sequences are drawn in solid for [Fe/H] $= 0$, dashed for [Fe/H] $=
-1$, and dotted for [Fe/H] $= -2$. The color codes for different
authors are noted in the figure (``Lejeune''is \citet{lej98},
``Green'' is the updated \citet{green88} table, ``V\&C'' is
\citet{vc03}), and ``Empirical'' refers to this work.
\label{fig:bvvi.cool}
}
\end{figure}

One region of parameter space
that is of keen interest is that of low stellar temperatures. We check
our results for cool stars against \citet{vc03}, \citet{lej98}, an
update of the \citet{green88} color table used in the Yonsei-Yale
isochrones \citep{yi01}, and, for good measure, the synthetic colors
of \citet{wor94b} in Figure \ref{fig:bvvi.cool}. The coolest giants
are very important for integrated-light studies of spectral features
such as TiO that become strong only in these stars, and for surface
brightness fluctuation (SBF) magnitudes, especially at red colors, that
depend on these stars because of the $L^2$ dependence of an SBF
magnitude. 

In Figure \ref{fig:bvvi.cool} our fits are shown as black lines. They
were fitted to the $B-V$ - $V-K$ and $V-I$ - $V-K$ diagrams, so it is
no surprise that they still fit in this color-color plane. The
updated-Green calibration follows an extrapolation of the giant
sequence off into regions not occupied by stars, while the dwarf
sequence for solar abundance follows the stars very well. There is
considerable metallicity dependence in the Green calibration that the
stars do not appear to share. The \citet{vc03} sequences follow both
dwarfs and giants fairly well, with a fairly good (small) abundance
dependence. The oscillations in the solar metallicity giant track are
a reflection of actual values in their data tables. The coolest
temperature reached by \citet{vc03} is 3000 K. The \citet{lej98}
calibration is based on corrected synthetic fluxes. In this case, the
dwarfs and giants track together with little or no gravity separation
until, at a temperature well within the tabulated range of
applicability, the values become wild.

\subsection{Colors not Explicity Fit}

\begin{figure}

\plotone{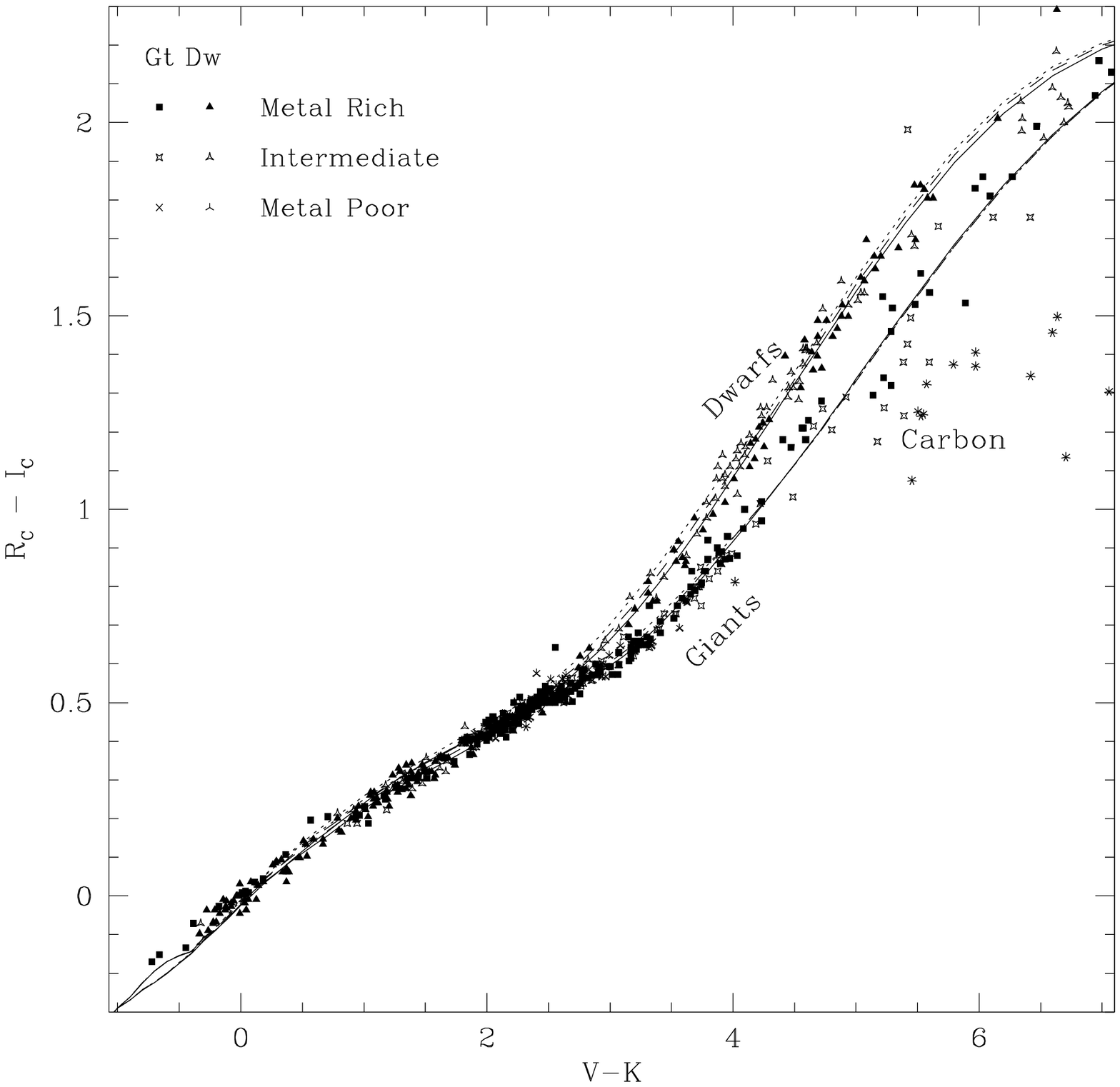}
\caption{$R-I$ is plotted as a function of $V-K$. 
Stars of different giant/dwarf status and abundance are plotted with
different symbols according to the key.
Stars with [Fe/H] $>
-0.2$ are considered metal rich, stars with [Fe/H] $< -1.2$ are
considered metal poor, and stars between these values are considered
intermediate in metallicity. Lines are coded as in Figure
\ref{fig:bvvi.cool}. This color was not fit during the calibration
process. 
\label{fig:rivk}
}
\end{figure}

Besides author comparisons, another way to check our results is to
plot colors that were not fitted explicitly to see if the implicit
color dependences are correctly modeled.  $R-I$ is one such, and is
illustrated in Figure \ref{fig:rivk}. For this color, the fits were
versus $V-R$ and $V-I$, for slightly different samples of stars. The
$R-I$ fitted tracks fall among the stars fairly well, except for a
hard-to-see reversal around $V-K=1.5$ where the giants become $\approx 0.02$
mag redder than the dwarfs. This 0.02 mag shift is probably incorrect,
but it gives a valuable indication of the reliability of the
color-color fits.


\subsection{The K Dwarf Desert}

\begin{figure}
\plotone{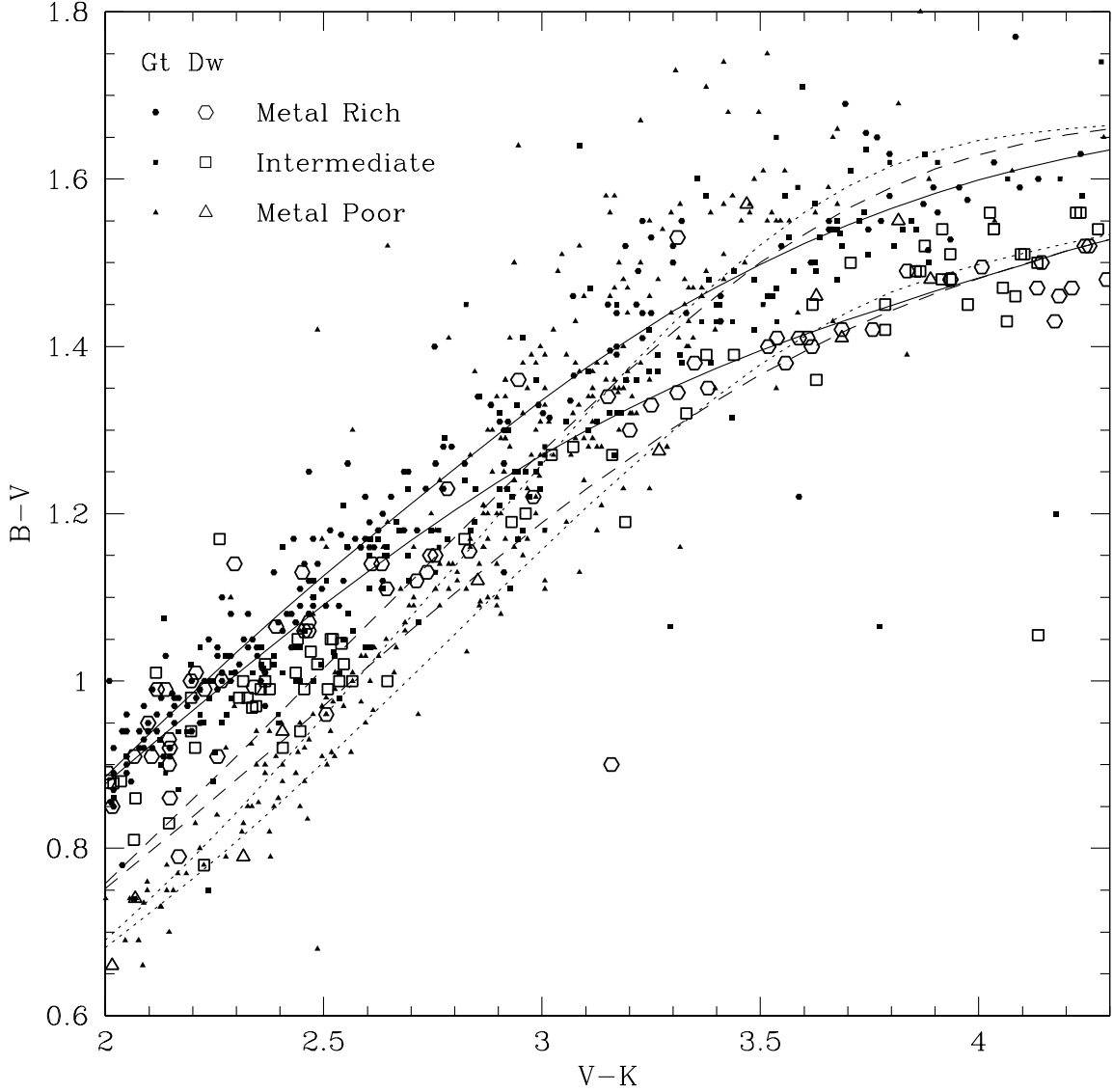}
\caption{The figure show a small section of the $B-V$, $V-K$
color-color diagram. Dwarfs are drawn with larger symbols than giants
for emphasis according to the key in the figure.  Stars with [Fe/H] $>
-0.2$ are considered metal rich, stars with [Fe/H] $< -1.2$ are
considered metal poor, and stars between these values are considered
intermediate in metallicity.  Calibrations for giant and dwarf
color-color sequences are drawn in solid for [Fe/H] $= 0$, dashed for
[Fe/H] $= -1$, and dotted for [Fe/H] $= -2$. At red color, the dwarfs
follow the bluer $B-V$ tracks. This is a region of uncertainty, as
discussed in the text.
\label{fig:kdesert}
}
\end{figure}

Reliability must be a function of temperature regime. One particular
troublesome area is that of K-type dwarfs and the damping of the
abundance sensitivity going toward cool stars. In Figure
\ref{fig:kdesert}, a $T_{eff}$ of 5000 K corresponds to $V-K \approx
2.2$ and $T_{eff} = 4000$ K corresponds to $V-K=3.4$. It is clear that
the magic combination of full photometry plus a good abundance
estimate is lacking from our data set for dwarfs in general and
metal-poor dwarfs in particular. Note the general lack of open symbols
(dwarfs) compared to filled (giants). This lack of data means that the
metallicity dependence of the dwarfs is inherited from the plentiful
giants in this temperature regime; an undesirable feature. At redder
colors, as their surface gravities diverge, the dwarfs and giants
separate in color. Simultaneously, the metallicity dependence appears
to reverse, at least in the giants. It is not completely clear from
the present data what should be happening with the dwarfs, although
they do seem to mirror the giants. The polynomials do their best to
smoothly flow through all of this, but we judge it unlikely that they
have truly captured the essence of the color behavior in this regime,
as it is not clear to our eyes exactly what should be happening (it
seems likely that some of the photometry is bad). The color reversal
with [Fe/H] is an issue only for $U-B$ and $B-V$ colors, although the
paucity of K dwarf data is of some concern for all colors, as the
metallicity dependence is relatively unconstrained.


\subsection{Error Propagation}

The principle source of error in the color-color fits is finding a
suitable polynomial to follow the various twists and turns that the
colors take. We fit the colors in five segments, with
multiply-redundant overlap in color, and used the overlap regions to
estimate the error from polynomial fitting. With typically hundreds of
stars available for each fit, random photometric uncertainty is not a
concern (though, of course, systematic uncertainty is). The median fit
uncertainty over all temperatures, gravities, and abundances is listed
in Table \ref{tab1}. We also thought it useful to propagate errors in
the final subroutine so that uncertainties in the effective
temperature scale could be translated to uncertainties in color. For
this we used the various $T_{eff}$ relations plotted in Figures
\ref{fig:mash2gt} and \ref{fig:mash2dw} and a couple of others to
roughly estimate a percentage error as a function of temperature. This
is given in Table \ref{tab2}. Note that the errors in Table \ref{tab2}
for cool stars are more applicable to giants than dwarfs; dwarf
temperatures seem more uncertain than those of giants, but we didn't
have enough dwarf calibrations to estimate this very well, so we left
it alone. For color $I$ with color error $\sigma_I$ and a temperature
error $\sigma_T$, errors propagate in the elementary way:

\begin{equation}
\sigma^2 = \sigma_I^2 + \bigl( {{{\rm d} I}\over{{\rm d} T}}\sigma_T    \bigr)^2 .
\end{equation}

\begin{deluxetable}{lr}
\tablecaption{Median Polynomial Fit Uncertainty \label{tab1}}
\tablewidth{0pt}
\tablehead{\colhead{Color} & \colhead{$\sigma$ (mag)} }
\startdata
$U-B$ & 0.071 \\
$B-V$ & 0.017 \\
$V-R$ & 0.010 \\
$V-I$ & 0.011 \\
$J-K$ & 0.004 \\
$H-K$ & 0.002 \\
\enddata
\end{deluxetable}

\begin{deluxetable}{rr}
\tablecaption{Temperature Uncertainty Assumed \label{tab2}}
\tablewidth{0pt}
\tablehead{\colhead{$T_{eff}$ (K)} & \colhead{$\sigma$ (\%)} }
\startdata
50000 & 4.0 \\
20000 & 2.5 \\
10000 & 1.0 \\
6000 & 0.5 \\
4000 & 0.5 \\
3500 & 1.0 \\
3000 & 1.5 \\
2000 & 4.0 \\
\enddata
\end{deluxetable}

\subsection{Reddening Estimation Using M Dwarfs}

\begin{figure}
\plotone{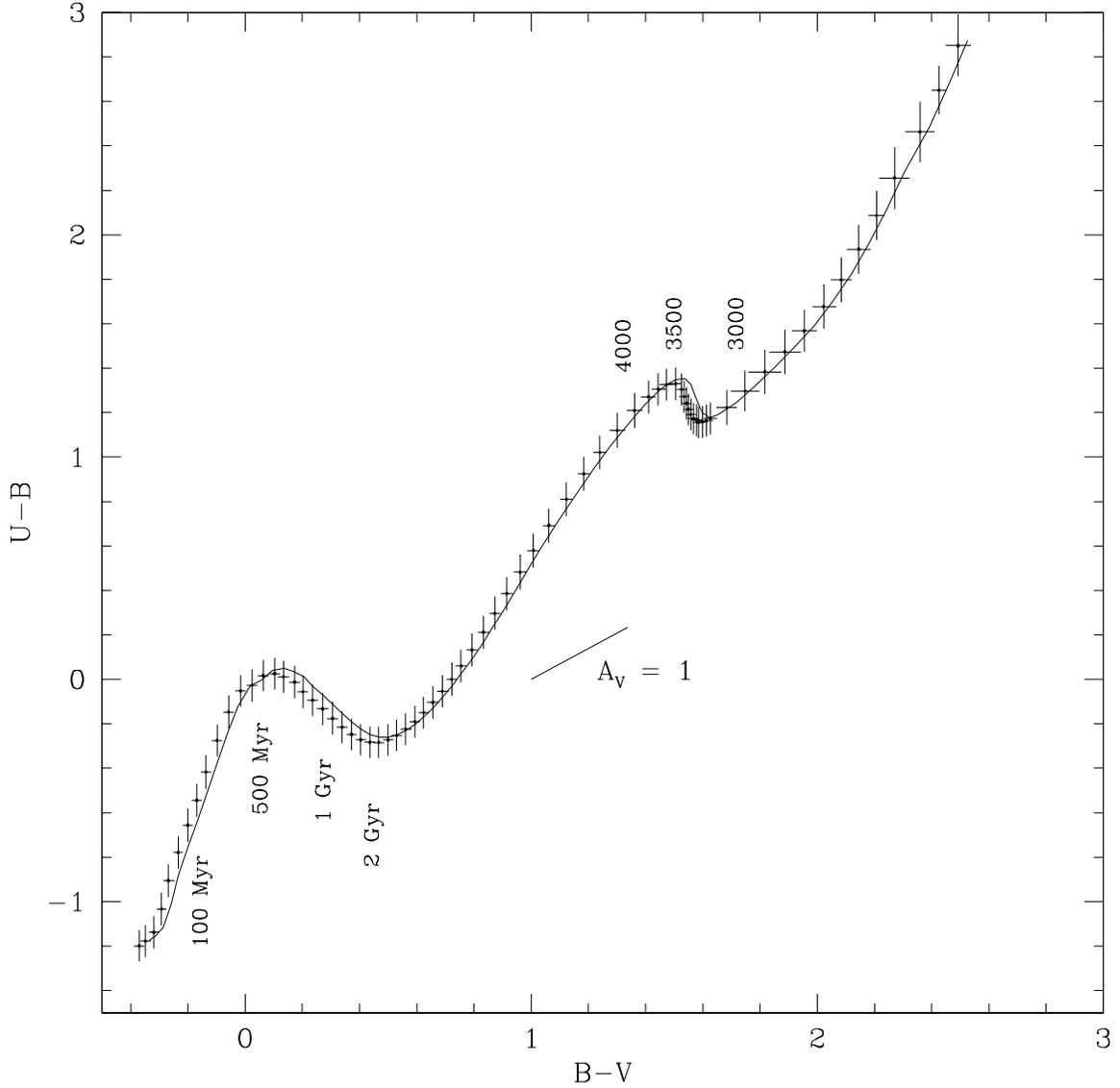}
\caption{The final color-color calibration for dwarfs only is shown in the $U-B$,
$B-V$ plane as dots with error bars attached, where the error bars include
fitting uncertainties and $T_{eff}$ uncertainties. An additional line
is shown that represents the color shift due to dust screening of $A_V
= 0.1$ mag. For illustrative purposes, a vector for $A_V = 1.0$ mag is
also sketched. The approximate blue limits of the bluest dwarfs at the
main sequence turnoffs of isochrones of various ages are marked with
the corresponding ages. Near the red bump feature, stellar effective
temperatures (degrees K) are indicated.
\label{fig:reds}
}
\end{figure}

Color-color diagrams have been used to derive a ``color excess'' from
which can be inferred a value for the dust extinction \citep{m53}. The
metallicity dependent color-color fits of this paper offer a general,
if not overly precise, method of generating a color-color plot for any
color combination as a function of abundance and gravity. The classic
$U-B$, $B-V$ diagram is shown in figure \ref{fig:reds} for dwarfs
only. The double inflection redward of zero color represents the rise
and fall of the Balmer break in B-type through A- and F-type stars. A
defect of this method is that it only works on clusters that have
A-type stars, that is, ones younger than about 1 Gyr that still have
dwarfs that hot. Interestingly, there is an additional color
inflection in the M dwarfs (cf. \citet{lej98}), roughly between 4000 K
and 3000 K, that may allow independent reddening estimates for old
clusters that have deep photometry. This inflection exists in almost
every color, although the $U$ band presents the most dramatic
manifestation of it.

The wiggle has been seen before in various colors and with variable
fidelity \citep{caldwell, bessell91, tapia, bryja} but with modern
telescopes and instrumentation, it may turn into an astrophysical
tool. It is caused by the onset of molecular absorption (TiO being the
number one culprit) across the M temperature range that radically
changes the underlying spectra shape, (c.f. \citet{bessell91} ).

If the $U$ band is utilized, the coolest stars involved have $U-I =
5.65$ mag according to our colors and $M_I=9.0$ according to diagrams
in \citet{legg92}. This leads to an absolute $M_U= 14.65$ mag. If a
modest telescope can reach $U=23$ as the KPNO 2.1-meter did in
\citet{kr95}, then the $U$ flux is readily detectable to about 500 pc
distance. For reference, the nearest ancient open cluster is M67 at
about 800 pc, so one would need a slightly larger telescope or better
$U$ sensitivity for $U$ to be useful. However, redder colors can also
be made to work at about the same confidence level relative to the
fitting errors.  The fitting errors are shown as error bars on points
in Fig. \ref{fig:reds}, as is another color sequence shifted by $A_V =
0.1$, shown as a line. We judge that this extinction is the smallest
that can be detected at all simply using the color-color fits we
present, and so is not particularly competitive with other methods as
it stands. Interestingly, at redder passbands, the M-type deflection
becomes less pronounced but the errors also decrease so that any
$A_\lambda$ extinction vector stays at about the same statistical
significance in most color-color planes. This does not solve the
problem, however, because observational measurement error becomes
larger than the fit error at $JHK$ wavelengths. Future refinements to
this reddening estimation method are possible and should be
encouraged.

Giants also show such an inflection. However, no Galactic cluster has
enough cool giants to populate the inflection region, globular cluster
giant branches being too warm, and open clusters being too low mass to
have many such giants. There may be limited application for local
group galaxy fields with resolved photometry; derivation of reddening
maps across the surface, for example. However, the compositeness of
the stellar populations of local galaxies may introduce too much error
in the scheme for it to be useful.

\subsection{Gravity Dependence Comparison}

\begin{figure}
\plotone{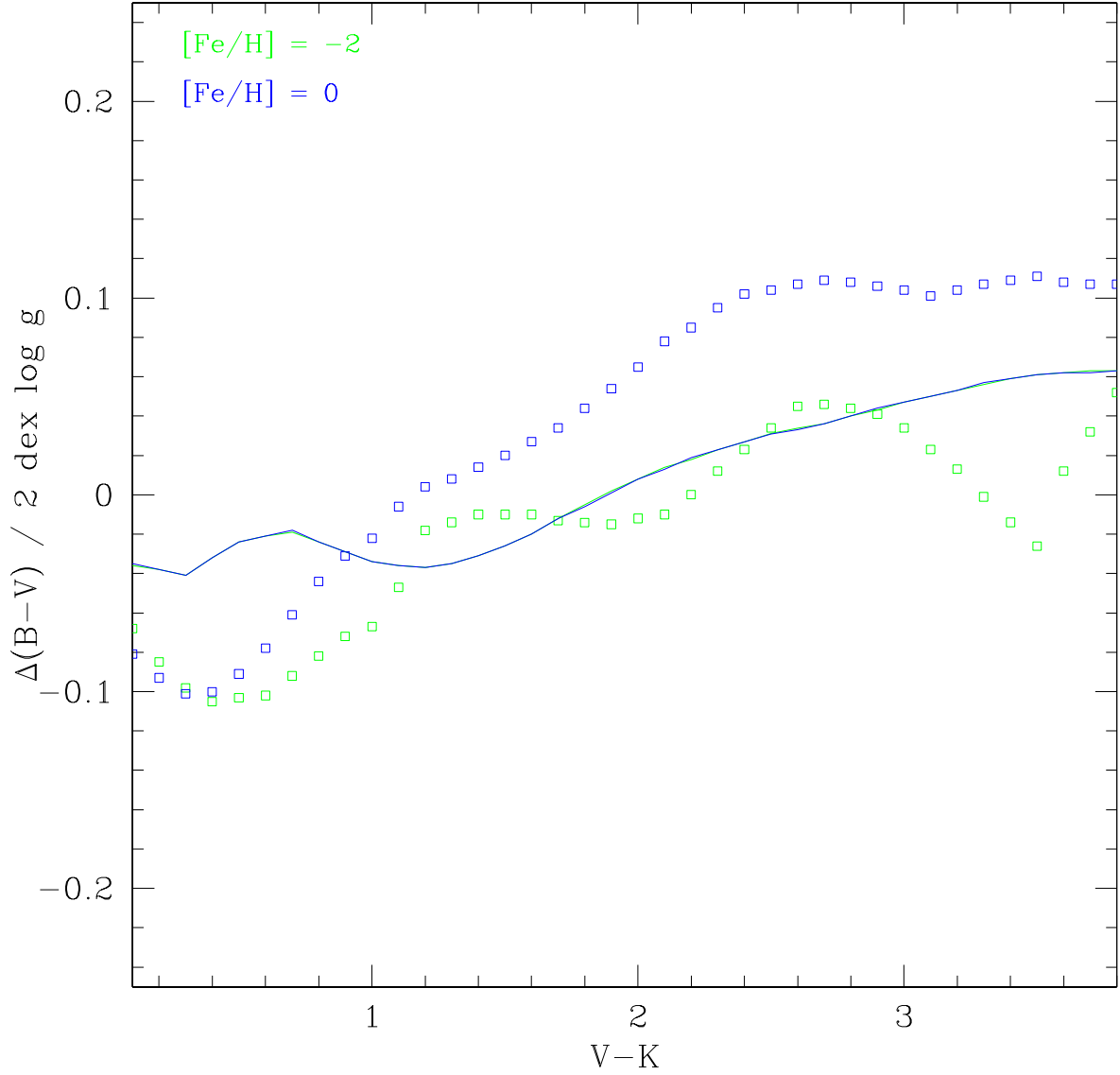}
\caption{ The change in $B-V$ color caused by a shift in log $g$ from
2 to 4, plotted against $V-K$ color. Green lines or symbols indicates
[Fe/H] $=-2$ and blue lines or symbols indicates [Fe/H] $=0$. Lines
are the present work, and both colors are present, but the lines
coincide because there was no crosstalk between [Fe/H] and log
$g$ in the color-color fitting process. Symbols are \citet{wor94b}.
\label{fig:diffcgbv}
}
\end{figure}

\begin{figure}
\plotone{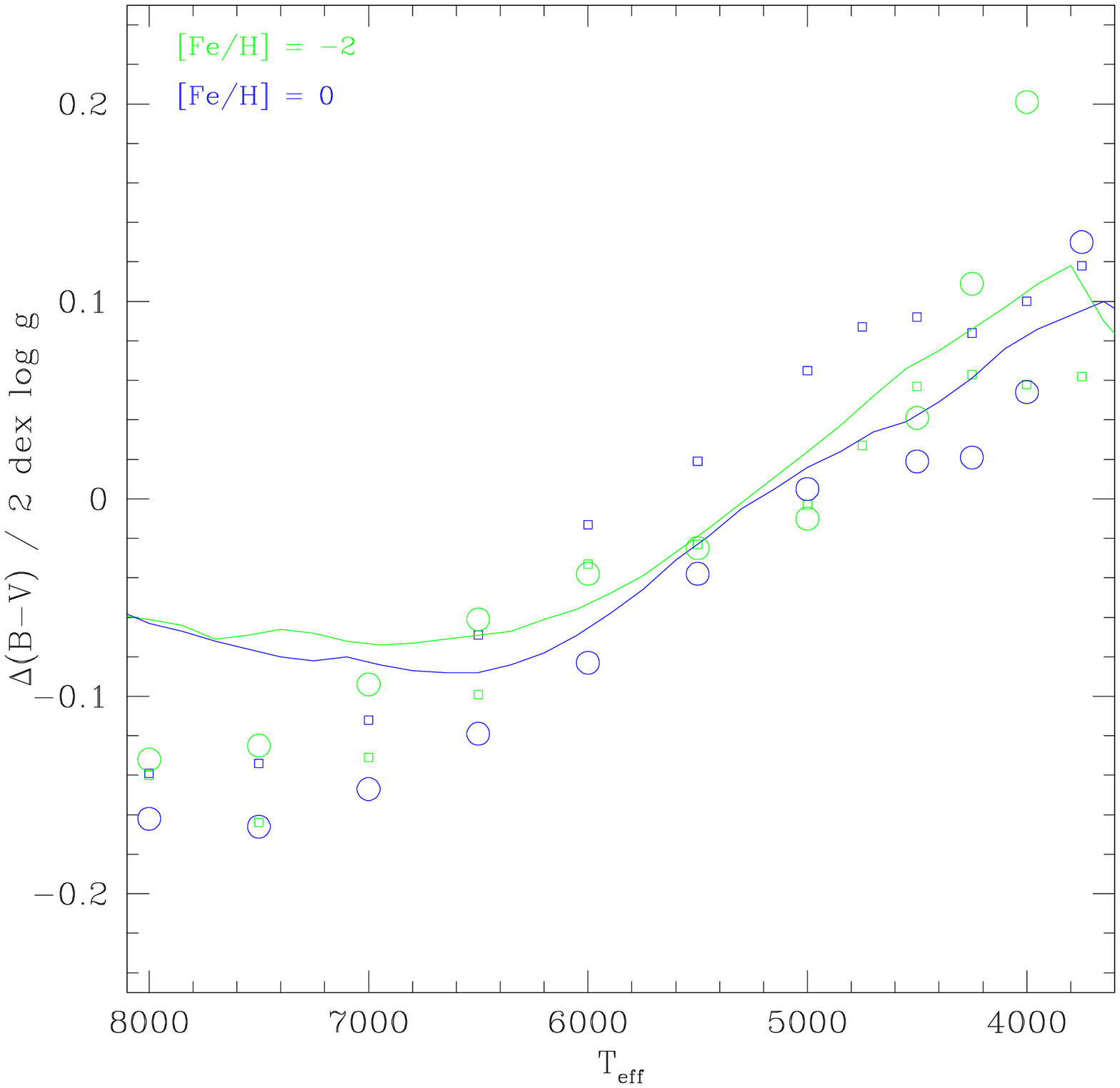}
\caption{
The change in $B-V$ color caused by a shift in log $g$ from
2 to 4, plotted against $T_{eff}$. Green lines or symbols indicates
[Fe/H] $=-2$ and blue lines or symbols indicates [Fe/H] $=0$. Lines
are the present work, small symbols are \citet{wor94b}, and large
symbols are the updated \citet{green88} table.
\label{fig:difgbv}
}
\end{figure}

The dimension of temperature is a downstream add-on component using
the method of this paper, but the dimensions of [Fe/H] and log $g$ are
inherited from the stellar catalog and can therefore be compared to
the predictions from previous calibrations in a fairly clean way. In
and near the M star temperature regime, we explicitly damped the
[Fe/H] dependence away, but the gravity dependence was freely fit. The
character of the data changes in that the dwarfs fork away from the
giants the cooler one goes, making any dependence more complicated
than linear rather suspicious. (No gravity dependence more than linear
was used in this work, in this regime.) By way of illustration, we
plot some color-derivatives for one color, $B-V$, with abundance held
fixed and gravity varied, in Figs. \ref{fig:diffcgbv} and
\ref{fig:difgbv}.

Fig. \ref{fig:diffcgbv} and Fig. \ref{fig:difgbv} show the same thing,
except for the $X$ axis choice. The $V-K$ color
(Fig. \ref{fig:diffcgbv}) was what was fit against, and only
calibrations that include both $B-V$ and $V-K$ can be
included. Fig. \ref{fig:difgbv} is plotted against $T_{eff}$ and can
be compared to more calibrations. In the latter figure, also, the
temperature scale difference cause the empirical trends to split. 

The conclusions from examining these and similar figures for many
colors are that the present work (1) resembles in gross other
calibrations, (2) tends show the smallest, mildest gravity dependence,
and (3) shows similar gravity dependence even at vastly different
metallicity regimes. All three of these conclusions appear to be
fairly robust, which should be a rather large concern, since the
delta-colors are quite substantial for most calibrations. An alarming example
of this, not illustrated, is $U-B$ for stars hotter than the sun, for
which the empirical (this work) gravity dependence is essentially
zero, but most other calibrations put it at $\Delta (B-V) / \Delta
({\rm log}\ g) \approx 0.15$ mag dex$^{-1}$.

\subsection{Future Temperature Scale Adjustments}

A topic beyond the scope of this paper deserves a comment, and that is
attachment of this calibration to existing theoretical stellar
evolutionary isochrones for purposes of comparing to star clusters and
for purposes of integrated light studies. As a test case, which we
intend to publish, multi-band photometry for two open clusters, M67
and NGC 6791, were collected from many sources and assembled into a
$UBVRIJHK$ data set. The color-color relations from these data sets
agrees well within expected errors with the color-color fits presented here.

\begin{figure}
\plotone{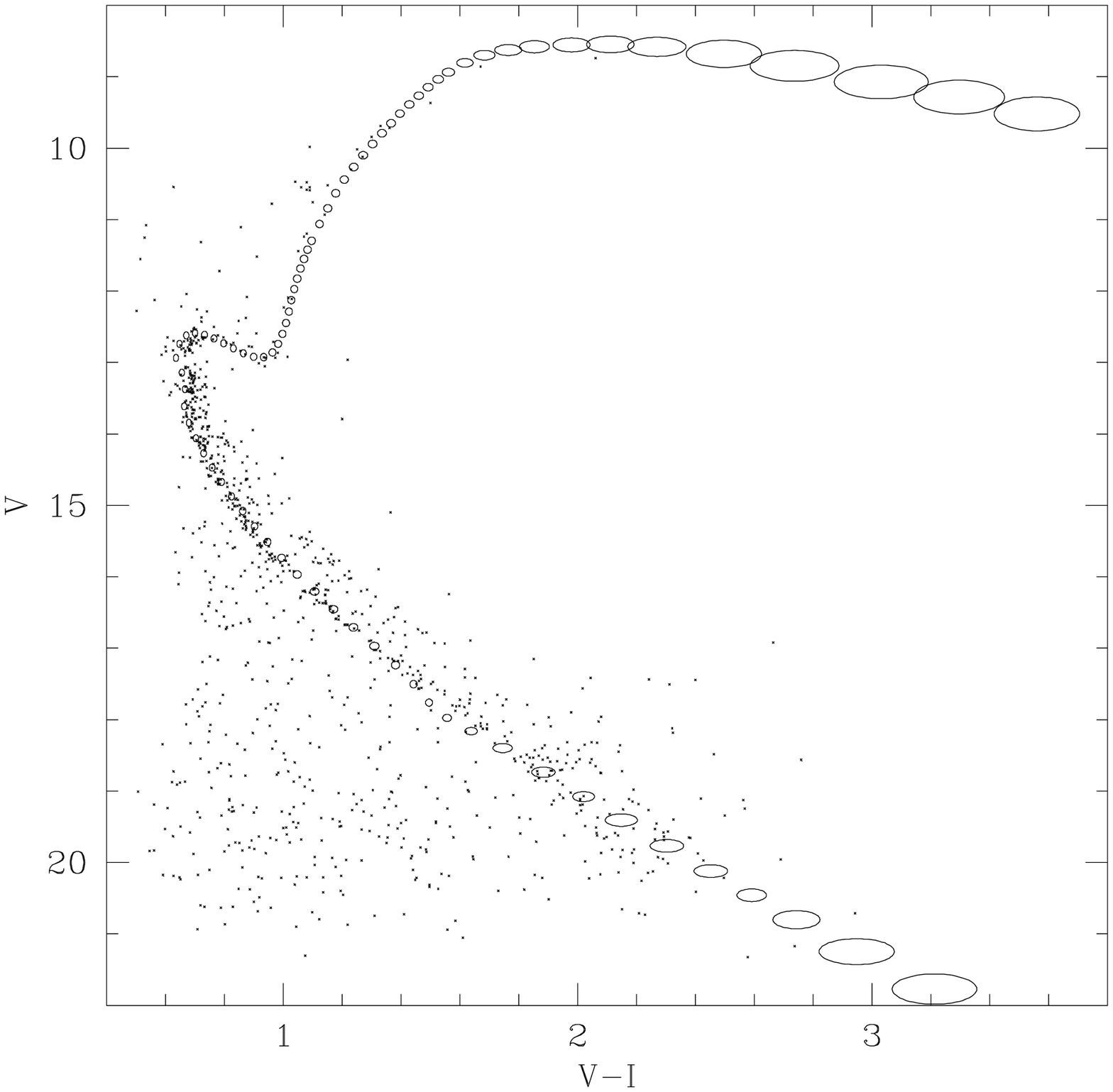}
\caption{
Color magnitude diagram and isochrones for open cluster M67. The
\citet{yi01} isochrones at solar metallicity and age 5 Gyr with the
present color calibration is shown as ellipses that represent the
propagated uncertainties. Distance modulus $(m-M)_V = 9.4$ and
reddening $E(V-I) = 0.02$ are assumed. The data is that of \citet{mont93}.
\label{fig:m67}
}
\end{figure}

However, the color-magnitude diagrams generated from \citet{yi01}
isochrones and this work, when compared to the real clusters, are not
so rosy. Figure \ref{fig:m67} shows a color magnitude diagram for open
cluster M67, along with ellipses that represent one-sigma errors on
our color calibration, and there are drifts between isochrone and data
that are substantially more than one sigma. Parenthetically, and with
an emphatic lack of surprise, one of the places of mismatch is the
late K dwarf region, among the temperatures where the empirical
calibrations are competing with the \citet{vc03} semiempirical
calibration. In that particular case, it is almost certainly the
attachment of the temperatures in our calibration that is causing the
wonkiness in the fit to the data.

In addition, for the finite set of data and models tried so far, a
fit is often satisfactorily only in one color. When $B-V$ and $V$ is fit, for
example, $V-K$ and $K$ do not fit for the same age and
reddening. Going into the realm of theoretical stellar models introduces
another layer of complexity that we are unable to cope with in this
paper, but it seems clear that the temperature scale attached to our
color-color relations is not, initially, going to mesh easily with
existing isochrone sets. We conjecture that the blame will be shared
between the temperature scale attached in this work, and the
temperature scales established in theory via mixing length theory, or
other convection prescriptions.


\section{Summary and Conclusion}

Johnson/Cousins photometry was combined with literature [Fe/H]
estimates to fit color-color diagrams as a function of gravity and
abundance. Literature-average temperature and bolometric correction
scales are attached to provide a global color-temperature relation for
stars with $-1.06 < V-K < 10.2 $. The $RI$ magnitudes are in the
Cousins system, and $JHK$ magnitudes are in the Bessell homogenized
system. The complete color-temperature table and a Fortran
interpolation program is available at http://astro.wsu.edu/models/. 

Several areas of improvement were noted in the main body of the paper,
including filling
photometry gaps, obtaining more accurate and on-system photometry,
knowing better log $g$ and [Fe/H] values, improving the statistics for
data-impoverished groups of stars such as K dwarfs, applying small
tweaks in the processing pipeline, and obtaining better empirical
temperature and bolometric correction relations, especially for
supergiants and M stars. 

A way to estimate dust extinction from M dwarf colors arises from an
inflection that exists in most colors relative to $V-K$. Unlike the
classic $UBV$ method, it can be used in old star clusters, but it does
not seem to promise much, if any, increase in accuracy for clusters
where both methods apply. The most sensitive band relative to
photometric error for the new extinction measure is the $U$ band, but
if the $U$ band is employed then clusters must be within a few hundred
parsecs for ground-based observatories to able to measure adequate $U$
fluxes.

\acknowledgements 

Major funding was provided by the National Science Foundation grants
AST-0307487, the New Standard Stellar Population Models project, and
AST-0346347. The SIMBAD data base, NASA's Astronomical Data Center,
and NASA's Astrophysics Data System were indispensible for this
project. GW would like to thank the undergraduates who have typed in
data pertaining to stars over the more than 14 years this project has
stretched. Brent Fisher \citep{wf96} and Joey Wroten at the University
of Michigan and Jared Lohr and Ben Norman at Washington State
University.


\end{document}